# Fabry-Perot resonance modes in a MoS$_2$-based vertical stacking cavity for strong light-matter coupling and topological phase singularity

Zhonglin Li[1,2], Yingying Wang[1,*], Xianglin Li[3,4], Bo Zhong[5], Wenjun Liu[1], and Zexiang Shen[6,*]

[1]Department of Optoelectronic Science, Harbin Institute of Technology at Weihai, Weihai 264209, China

[2]Department of physics, Harbin Institute of Technology, Harbin 150001, China

[3]Hunan First Normal University, No.1015, Fenglin Road (the 3rd), Yuelu District, Changsha, Hunan, China 410205

[4]Donguan NanoFrontier Microelectronics Equipment Co., Ltd, City of University Innovation, Songshan Lake, Guangdong, China

[5]School of Materials Science and Engineering, Harbin Institute of Technology at Weihai, Weihai 264209, China

[6]Division of Physics and Applied Physics, School of Physical and Mathematical Sciences, Nanyang Technological University, Singapore 637616, Singapore

a) Author to whom correspondences should be addressed:

yywang@hitwh.edu.cn; zexiang@ntu.edu.sg

**Abstract:**

Rich dielectric properties in atomic transition metal dichalcogenides (TMDs) enhance light-matter interactions and contribute to a variety of optical phenomena. The direct transfer of TMDs onto photonic crystals facilitates optical field confinement and modifies photon dispersion through the generation of polaritons. However, light-matter interaction is severely limited by this stacking method. This limitation can be significantly improved by constructing a vertical stacking cavity with alternating layers of dielectric material and monolayer $MoS_2$. This multilayer structure is proven to be a compact, versatile, and customizable platform for controlling Fabry-Perot cavity resonance mode. Angle-resolved reflectance further aids in studying resonance mode dispersion. Moreover, the strong light-matter interaction results in multiple perfect absorptions, with the monolayer $MoS_2$ significantly contributing to the absorption in this system, as schematically revealed by the electric field distribution. The multiple perfect absorptions produce an unusual amounts of phase singularities with topological pairs, whose generation, evolution, and annihilation can be controlled by adjusting cavity parameters. Our findings provide a flexible and consistent framework for optimizing light-matter interactions and supporting further studies on wavefront shaping, optical vortices, and topological variants.

## 1. Introduction

Two-dimensional (2D) transition metal dichalcogenides (TMDs) provide promising platforms for studying light-matter interactions due to their tunable band gap, large dielectric function, high excitonic binding energy, and strong oscillator strength at room temperature[1]. 2D-TMDs can be transferred onto arbitrary substrates without lattice-matching requirements, enabling the rational design of artificially stacked structures with tailored optical properties. Recently, the coupling of microcavity photons with excitons in 2D-TMDs has promoted the study of polariton physics including many-body quasiparticle interactions[2], coherent qubits for quantum communication[3,4], superfluidity, and Bose–Einstein condensation[5-7]. Various photonic structures, such as one-dimensional (1D) gratings and 2D photonic crystals with waveguide resonance modes, have been coupled to 2D-TMDs materials to achieve a variety of functionalities[8], including unity Fano reflection[8,9], far-field excitonic light emission[10,11] and the splitting of photon dispersions into polariton branches[12,13].

In the most commonly studied configurations, 2D-TMDs are directly transferred onto photonic crystals, where light propagates parallel to the atomic plane of the 2D-TMDs. 2D-TMDs function either by confining optical fields within photonic structures, modulating reflective/transmissive line shapes with Fano asymmetry, or splitting photon dispersion into polariton branches[13,14]. However, light-matter interaction is severely limited by this stacking method. This limitation could be significantly improved by creating a vertical cavity structure, where 2D-TMDs are positioned in the path of light propagation. The enhanced light-matter interaction facilitates the generation of zero light intensity, or local darkness, in a scalar field. The zero light intensity provides access to topological phase singularities, intensity control, polarization conservation, and filtering[15,16]. For instance, dimensional tuning of phase singularities from zero-dimensional (point) to 2D (sheet array) through phase gradient engineering enables arbitrary light intensity at the cross section [17, 18]. As a singular optics phenomenon, phase singularities using photochromic molecules or metamaterials allow for the development of highly sensitive optical sensors and the enhancement of light-matter coupling strength[19, 20].

Additionally, the study of resonance modes in atomic layer materials stacked within vertical Fabry-Perot (FP) cavities has been rarely explored, except for a recent study on superlattice structures based on monolayer 2D-TMDs and dielectric layers (h-BN and $Al_2O_3$), which exhibit unity absorption at excitonic resonance[21]. The investigation of resonance modes and topological phase singularities could promote the application of atomic layer materials in optical sensing, topological radiation, and topologically protected scattering.

In this work, by alternatively stacking dielectric layers and monolayer $MoS_2$ with different periods, a vertical FP cavity is constructed, which is demonstrated to be a compact, versatile, and customizable platform for controlling FP resonance modes. Angle-resolved reflectance clearly reveals the resonance mode dispersions both experimentally and theoretically. The resonance modes are further systematically explored by analyzing their electric field profiles. Additionally, the strong light-matter interaction facilitates the generation of multiple zero-reflection points, supports to the observation of topological phase singularities, and enables the generation, evolution, and annihilation of topological pairs. Our results demonstrate that TMD-based vertical cavities offer a promising platform for investigating topological photonics, as they allow precise tuning of topological charges and the generation of varying degrees of phase singularities.

## 2. Results and discussions

## 2.1 A multilayer vertical cavity facilitates the formation of FP resonance modes

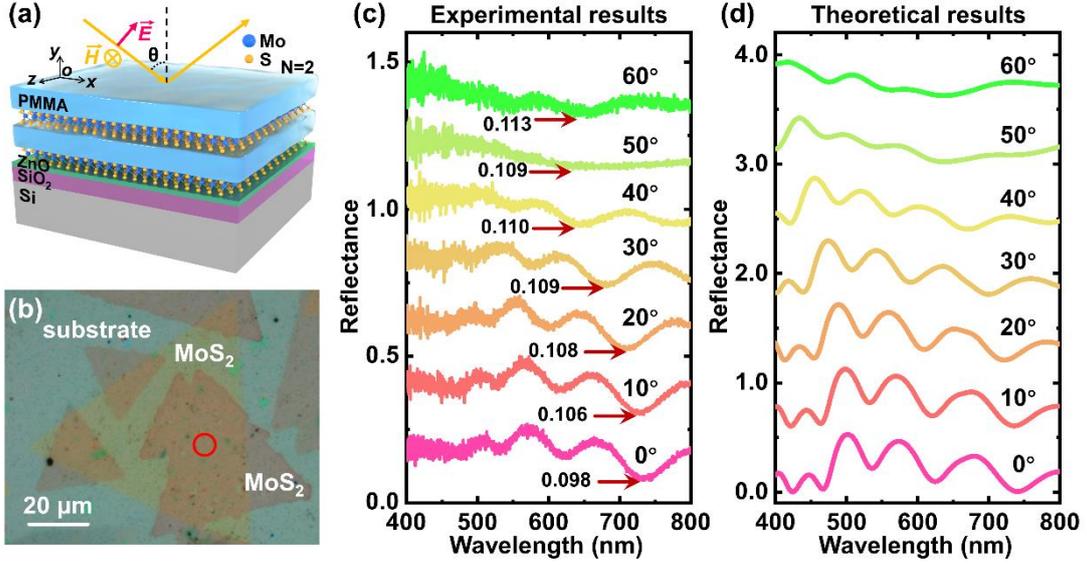

**Figure 1.** A multilayer vertical cavity facilitates the formation of resonance mode. (a) Schematic image of a multilayer vertical cavity composed of alternating unit cells of dielectric PMMA layers and monolayer $MoS_2$ with a period N=2. Light is incident on this cavity at an incidence angle $\theta$. The supported substrate consists of ZnO(34 nm)/$SiO_2$(300 nm)/Si. (b) Optical image of the multilayer cavity with N=2. (c,d) Experimental and theoretical results of angle-resolved reflectance for the N=2 cavity with p-polarized light.

The resonance modes of a vertical planar cavity can be interpreted using a basic three-layer structure, such as a free-standing thin film in air with a refractive index of n and a thickness of d, which supports FP resonance modes. FP resonance modes are standing waves formed by light wave propagating back and forth between two reflective interfaces. A reflection dip (destructive interference) occurs when light wave completes a round trip, and the following phase accumulation conditions are satisfied at a normal incidence,

$$\frac{2nd}{\lambda_0} + (\frac{\varphi_1 + \varphi_2}{2\pi}) = k + \frac{1}{2} \tag{1}$$

Here, $\varphi_1$ and $\varphi_2$ represent the interfacial phase shifts at the two interfaces, respectively, and $2nd/\lambda_0$ is the phase accumulation as light travels through the thin film. k is the order of resonance (k=0, 1, 2, 3…), and $\lambda_0$ is the wavelength at the reflection dip at normal incidence. When light is incident at an angle $\theta_i$, as schematically illustrated in Figure 1a, the cavity photon dispersion for a free standing film-based cavity following Eq.(2) [22].

$$\lambda(\theta_i) \approx \lambda_0 \sqrt{1 - \frac{\sin^2(\theta_i)}{n^2}} \quad (2)$$

$\lambda(\theta_i)$ is wavelength at the reflection dip for different incident angles. For an FP vertical cavity constructed by more than three layers, multiple interfacial phase shifts and phase accumulations occur in each individual thin film. The total phase accumulation for the reflection dips is also described by Eq.(1).

Figure 1a shows a schematic image of an FP cavity constructed by alternately stacking unit cells of dielectric Polymethyl methacrylate (PMMA) layer and monolayer MoS$_2$ with a period N=2. The substrate consists of ZnO(34 nm)/SiO$_2$(300 nm)/Si. Unlike conventional distributed Bragg reflectors, which consist of multiple dielectric layers with alternating high and low refractive indices with an optical path length of $\lambda/4$, the ZnO/SiO$_2$/Si substrate functions as an effective reflector for coupling cavity photons with excitons in 2D materials due to interference effect[23]. The surface of the 2D materials is defined as the x-z plane, with the y-direction perpendicular to the atomic plane of the 2D materials. For light with different polarizations, p-polarized light is defined as transverse-magnetic (TM) mode, while s-polarized light is defined as transverse-electric (TE) mode. The results for TE mode are provided in Supplementary Notes 1 and 2.

Monolayer MoS$_2$ is grown using the chemical vapor deposition (CVD) method on a SiO$_2$/Si substrate, and three-point star, and multiple-point star shapes of MoS$_2$ are observed during growth. A PMMA film is then spin-coated onto the MoS$_2$. The unit cell of the PMMA layer and monolayer MoS$_2$ is prepared for transfer, making the structure compact and versatile for integrated devices (Method). By repeatedly transferring the unit cell, a vertical van der Waals planar photonic structure with varying periods can be fabricated, and the alternating refractive indices in the vertical direction provide extensive tuning of the resonance modes.

Figure 1b shows the optical image of the N=2 FP cavity. The thickness of the PMMA layer is 500 nm. The sequential stacking of monolayer MoS$_2$ results in a twisted pattern, as seen in this figure, demonstrating the successful fabrication of a periodic structure. The PMMA layer (500 nm) is thick enough to prevent interactions between adjacent monolayer MoS$_2$, allowing it to retains its monolayer electronic properties when integrated into a multilayer stack[21]. Figure 1c gives the experimental and theoretical results of angle-resolved reflectance for the N=2 cavity in TM mode. The experimental results were taken from the position highlighted by a red circle in Figure 1b and closely match the theoretical calculations (Method). The theoretical results were calculated using the transfer matrix method. The low reflectance around 400 nm is due to the weak light intensity of the LED source configured in the spectrometer for this range. As seen in Figures 1c and 1d, for a fixed incident angle, multiple extremely low reflectance points (reflection dips) in the visible light range are observed, indicating the generation of FP resonance modes. These dips exhibit a blue shift as the incident angle increases due to cavity photon dispersion. The lowest reflectance values at different incident angles are individually marked and shown in the figure.

**2.2. The number of resonance mode could be tuned by the structure parameter**

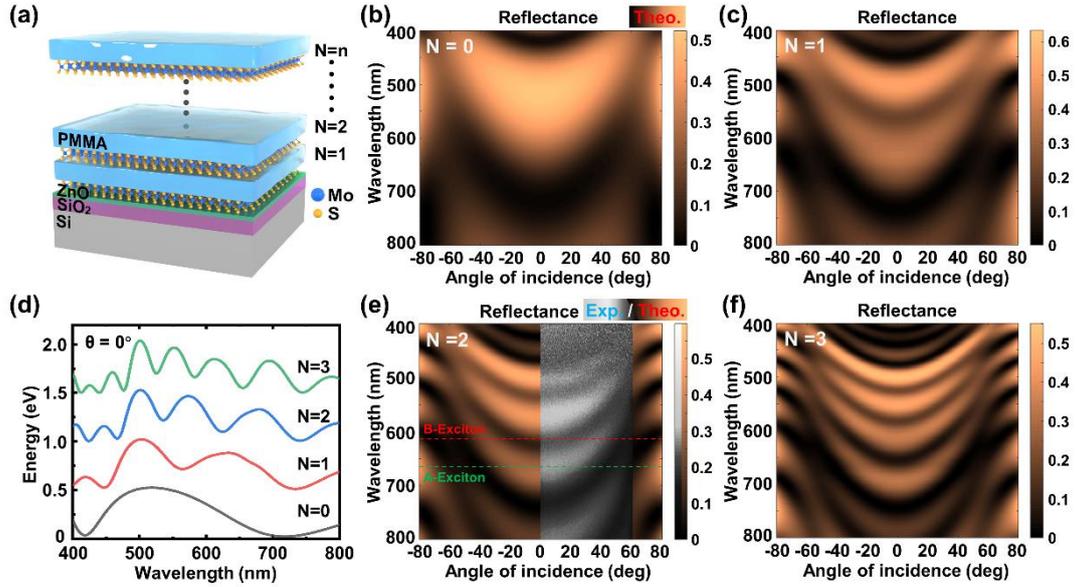

**Figure 2.** The number of resonance modes can be tuned by the structure parameters. (a) Schematic image of a multilayer cavity constructed with different periods N. (b-f) The angle-resolved reflectance of the vertical cavity with N increasing from 0 to 3 for TM modes, shown theoretically. The experimental observations for the N=2 cavity are provided and compared in Figure 2e, which closely match the theoretical observations. (d) The reflectance spectrum for the cavity with different periods N at an incident angle of 0°.

Figure 2a presents a schematic image of the vertical multilayer cavity structure with different period N, which supports FP resonance modes. Figures 2b-f give the theoretical results of angle-resolved reflectance for cavities with N increasing from 0 to 3 for TM modes, providing the same information as $k_∥$ space spectroscopy. Here, $k_∥$ is the in-plane (x-z plane) component of incident wavevector, given by $k_∥ = k\sin\theta$, with k as the incident wavevector and $\theta$ as the incident angle[22]. The experimental results for the N=2 cavity are summarized in Figure 2e, which closely match the theoretical observations (method). The A and B exciton absorption in $MoS_2$ are highlighted by colored dashed lines in Figure 2e. The theoretical results for the N=2 cavity for TE modes are provided in Supplementary Notes 1 and 2. From these figures, three key observations can be made: First, the resonance mode blue shifts as the incident angle increases, following the dispersion of the cavity photon. Second, as N increases, the number of resonance modes increases significantly. Third, the cavity photon of different resonance modes exhibits parabolic dispersions. As the photon energy increases, the dispersion changes from much steeper to much shallower. Figure 2d shows the reflectance spectrum for the cavity with different periods N at an incident angle of 0°.

**2.3. The resonance modes of the FP cavity are clearly distinguished and identified**

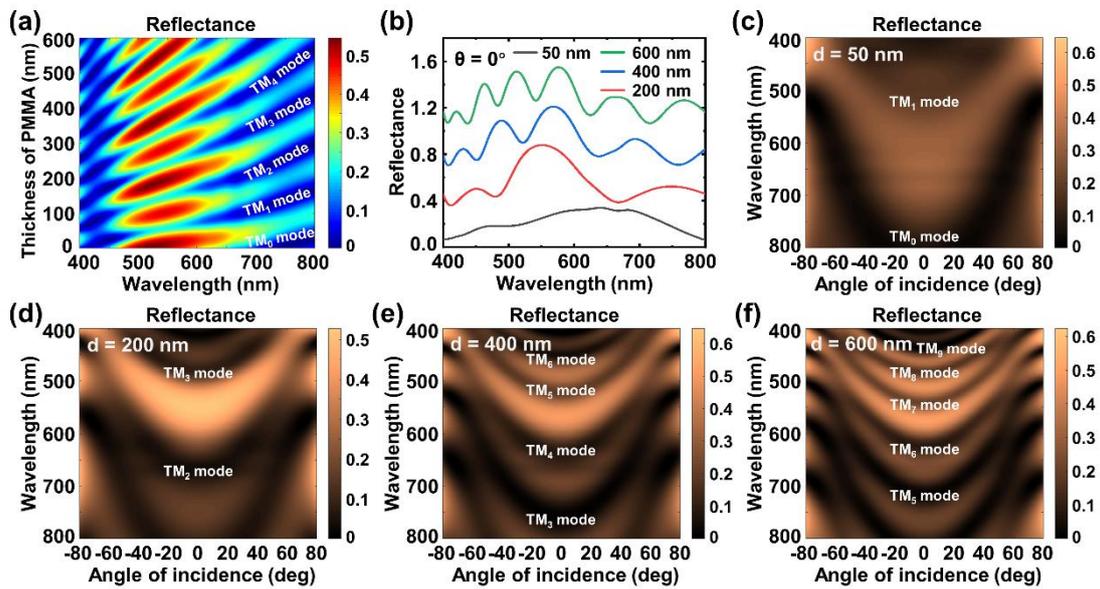

**Figure 3.** The resonance modes of the FP cavity are clearly distinguished and identified. (a) The theoretical reflectance mapping of an N=2 cavity in the space of wavelength and PMMA thickness at normal incidence. The resonance mode is clearly identified as the PMMA thickness increases. (b) The theoretical reflectance of the N=2 cavity at different PMMA thicknesses under normal incidence. (c-f) The theoretical reflectance of the N=2 cavity in the space of incident angle and wavelength for PMMA with different thicknesses in TM modes. The resonance modes are distinguished and marked consecutively.

Figure 3a shows the theoretical reflectance of an N=2 cavity in terms of PMMA thickness and wavelength at normal incidence. The thickness of the two-layer PMMA is tuned simultaneously. It can be seen that as the PMMA thickness increases, high-order FP resonance modes gradually appeared. These resonance modes are defined as $TM_k$ ($TE_k$) for p-polarized (s-polarized) light, with k determined by the order of occurrence. The fundamental mode and higher-order FP resonance modes are consecutively identified as $TM_0$, $TM_1$, $TM_2$, and so on. Figure 3b shows the reflectance of the N=2 cavity at different PMMA thicknesses at normal incidence. As the PMMA thickness increases, the number of resonance modes significantly increase. Figures 3c-3f show the theoretical reflectance of the N=2 cavity for PMMA of different thicknesses in terms of incident angle and wavelength for the TM mode. For 50 nm PMMA, the fundamental and first-order resonance modes are observed, and as the PMMA thickness increases, multiple resonance modes are generated, which are clearly identified and distinguished. Therefore, the PMMA thickness promotes the tuning of resonance modes.

**2.4. Intensity distributions at zero-reflectance for N=2 cavity**

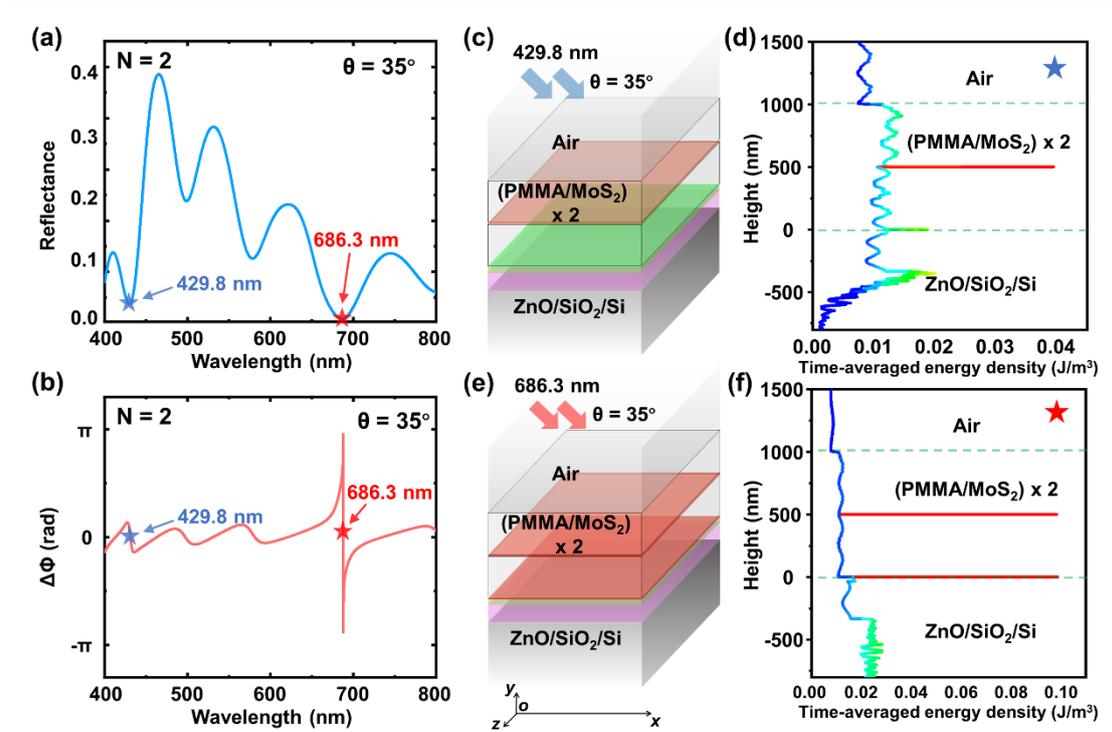

**Figure 4.** Light intensity distributions at zero reflectance at 429.8 nm and 686.3 nm. (a) Reflectance of the N=2 cavity at an incident angle of 35º for the TM mode. (b) The phase difference between p-polarized and s-polarized light of the N=2 cavity at an incident angle of 35º for the TM mode. (c,e) Schematic images of N=2 cavity structures at different wavelengths. (d,f) Time-averaged energy density distribution for 429.8 nm and 686.3 nm.

When light propagates along the y-axis, and is confined to the xz plane, the light intensity distributions for cavity photon of different wavelengths are thoroughly investigated. Figure 4a shows the reflectance of the N=2 cavity at an incident angle of 35º for the TM mode. Figure 4b gives the phase difference between p-polarized and s-polarized light of the N=2 cavity at an incident angle of 35º for the TM mode. Zero reflectance and phase jumps at 429.8 nm and 686.3 nm are observed, highlighted by blue and red stars in Figures 4a and 4b, respectively.

Figures 4c and 4e show schematic images of this van der Waals structure with N=2, where the positions of $MoS_2$ are highlighted by colored planes, spaced by transparent 500 nm PMMA. The surface of ZnO is defined as 0 along the y-axis. Figures 4d and 4f give the time-averaged energy density distribution for the 429.8 nm and 686.3 nm along y-axis, individually. For 686.3 nm, the energy of the incident light is confined within the $MoS_2$ layers with only a minor change in intensity. In contrast, for 429.8 nm, the energy of the incident light is also confined within the $MoS_2$ layers but decays exponentially. Consequently, at zero reflectance for different wavelengths, the energy of the electric field is localized in the monolayer $MoS_2$, which confines the incident light and results in zero reflectance. Despite $MoS_2$ having only monolayer thickness, the significant absorption contributes to the generation of zero reflectance. In addition, the reflectance mapping for the N=8 cavity in terms of incident angle and wavelength, and the electric field distributions at zero reflectance for p-polarized light (TM mode) and s-polarized light (TE mode) are provided in the supplementary notes 3 and 4.

**2.5. Strong light-matter interactions result in the formation of topological phase**

singularities

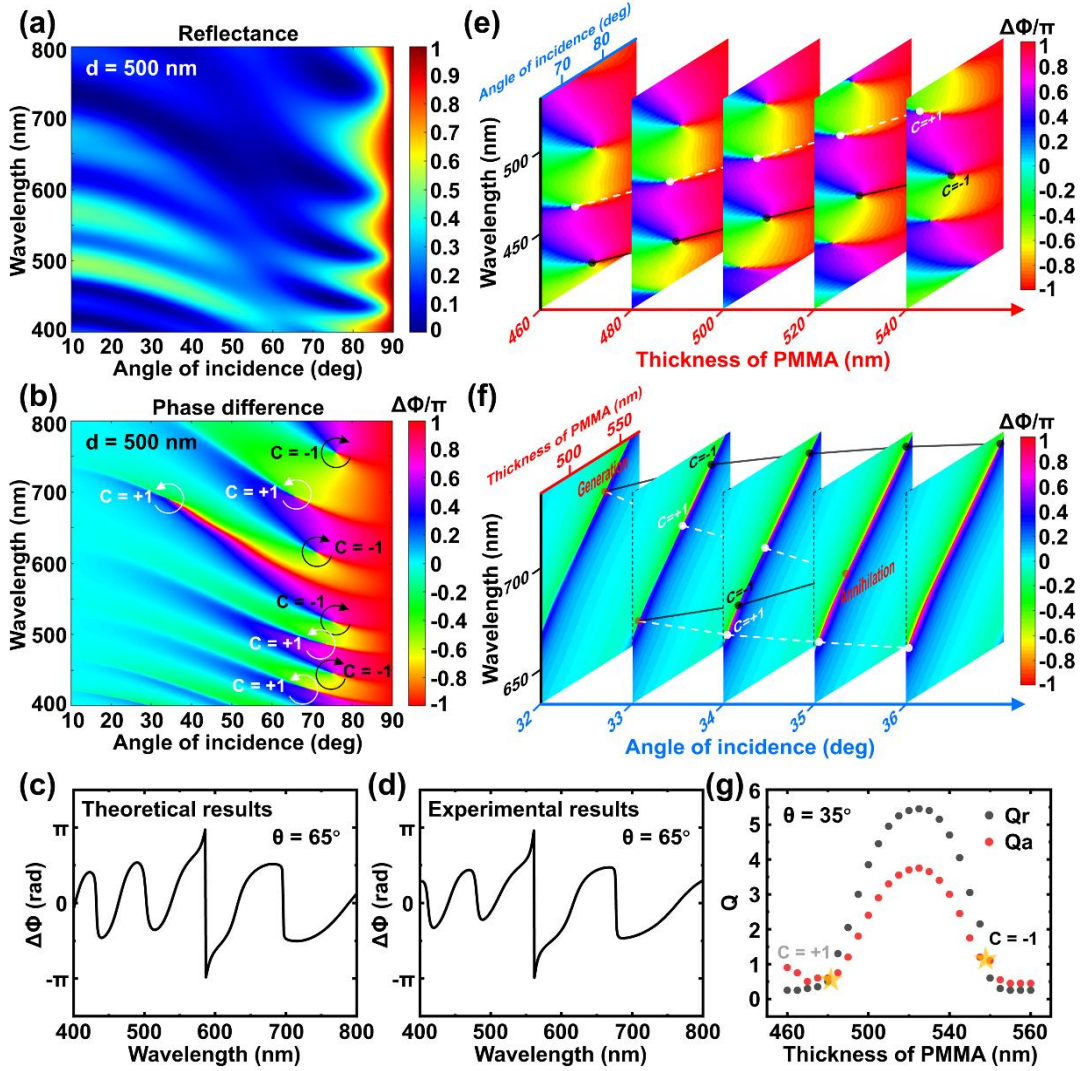

**Figure 5.** Strong light-matter interaction results in the formation of topological phase singularities. (a) Theoretical reflectance for an N=2 cavity in terms of incident angle and wavelength. The thickness of the PMMA layer is 500 nm. In this figure, the reflectance is calculated with p-polarized light when light is incident at a specific angle. (b) The phase difference between p-polarized and s-polarized light in terms of incident angle and wavelength for the N=2 cavity. Multiple phase singularities associated with topological charges are observed and marked. (c,d) The phase difference of different polarized light at an incident angle of 65° is shown both theoretically and experimentally for the N=2 cavity. (e) The position of the topological charge pair (C=±1) can be tuned by PMMA thickness, as demonstrated in the map of phase difference in terms of incident angle and wavelength. (f) The generation, evolution, and annihilation of topological charge pairs can be tuned by the incident angle, as shown in the map of phase difference in terms of PMMA thickness and wavelength. (g) The competition between radiation rate and absorption rate determines the position or nature of the topological.

Perfect absorption using 2D-TMDs materials at an ultrathin limit is a highly sought-after regime of light-matter interaction, which can be enhanced by angle-resolved excitation[24, 25]. Here, it is found the angle-resolved enhanced light-matter interaction in a monolayer $MoS_2$ based vertical cavity facilitates the generation of zero reflectance and topological phase singularities. Figure 5a gives the theoretical results of the angle-resolved reflectance of the N=2 cavity in terms of wavelength and incident angle. The thickness of the PMMA layer is 500 nm. In this figure, the reflectance is calculated with p-polarized light when light is incident at a specific angle, using the transfer matrix method. Multiple zero-reflectance points are observed, which originate from the strong light absorption by two isolated monolayer $MoS_2$ in this cavity. Figure 5b shows the theoretical map of the reflection phase difference $\Delta\Phi$ between p-polarized and s-polarized light in terms of incident angle and wavelength, when polarized light is reflected from the cavity surface. As seen in this figure, numerous phase singularities (vortices) are observed with an undefined phase at their centers, and around which a round trip accumulation of $\pm 2\pi$ phase are observed.

Topological charges C=±1 are used to quantitatively describe the phase singularities, which are marked by white and black circles in Figure 5b. The topological charge C is defined as $C = \frac{1}{2\pi} \oint_\gamma \mathrm{grad}(\varphi) ds$, whose value is obtained by integrating the phase gradient along a path enclosing the phase singularity and divided the result by $2\pi$. Besides, it is found that the positions of the phase singularities correspond to those of zero reflectance in Figure 5a and Figure S5c, indicating a loss-assisted phase modulation mechanism[26-28]. Furthermore, these unusual phase differences are typically generated in pairs, as shown in Figure 5b. The reflectance mapping for the N=2 cavity, considering s-polarized and p-polarized light at the given incident angle and wavelength, along with the phase of s-polarized and p-polarized light, can be found in supplementary notes 5. Figures 5c and 5d give the theoretical and experimental results of the phase difference at an incident angle of 65° for the N=2 cavity. The experimental results, measured by a ME-L-L ellipsometer, closely reproduce the theoretical results (Method). A $2\pi$ phase jump at the wavelength of 561 nm, and with increasing incident angle, multiple large phase jumps at different wavelengths are observed, which enables the fabrication of multiple-channel sensors with high throughput.

In addition, the generation, evolution, and annihilation of topological pairs are further explored in Figures 5e and 5f. The theoretical phase difference $\Delta\Phi$ in terms of wavelength and incident angle for the N=2 cavity at different PMMA thicknesses is given in Figure 5e. The thickness of the two-layer PMMA is tuned simultaneously. For 460 nm PMMA, a pair of topological charges C=±1 is observed, marked by white and black dots, respectively. As the PMMA thickness increases, the movement of the topological charges in spectral space is observed. The white dashed line indicates the evolution trajectory for C=+1, while the black line shows the trajectory for C=-1. Consequently, the structure parameter, such as PMMA thickness, facilitates the tuning of topological charge positions.

In addition, the generation, evolution, and annihilation of topological charge pairs are further studied in Figure 5f, which shows the phase difference $\Delta\Phi$ in terms of wavelength and PMMA thickness for the N=2 cavity at different incident angles. For an incident angle of 32°, a phase singularity (marked as a red dot) is observed. This singularity point can be considered as a singularity nucleus, which eventually develops into a topological charge pair as the incident angle increases. A pair of singularities, C=±1, at wavelengths of 707.9 nm and 729.2 nm evolves from this singularity nucleus, as seen at 33º. Additionally, another singularity nucleus is generated (also marked as a red dot), which similarly evolves into a topological pair with an increase in the incident angle. The evolutions of the topological charges, C=-1 and C=+1, are marked by black lines and white dashed lines, individually. At an incident angle of 35º, the approach of C=-1 and the adjacent C=+1 results in the annihilation of the topological pair, marked by a red dot. At an incident angle of 36º, the annihilation process is fully completed, with the topological charges reaching the boundary in spectral space, as shown in this figure. Therefore, by adjusting structure parameters such as incident angle, PMMA thickness, and wavelength, we can monitor the nucleation, generation, evolution, and annihilation of topological charge pairs in spectral space.

It is further found that the positive or negative value of the topological charge is related to the competition between the radiation process and the absorption process in this cavity. For a vertical stack supported by thick silicon, which attenuates any transmitted light, this system can be defined as a one-port single-mode resonator. The complex reflection coefficient of light reflected from the cavity surface around the resonance frequency ($\omega_0$) can be calculated by Eq.(3), which gives the amount of phase shift of the incident light upon reflection, [29-31]

$$r = -1 + \frac{1/Q_r}{-i(\omega/\omega_0 - 1) + (1/Q_r + 1/Q_a)/2} \qquad (3)$$

Here, $\omega$ and $\omega_0$ are the frequencies of the incident light around and at resonance, individually. This resonator is defined by two parameters, $Q_r$ and $Q_a$, which are relate to radiation rate $\gamma_r$ and absorption rate via $\gamma_r = \omega_0/2Q_r$ and $\gamma_a = \omega_0 / 2Q_a$, individually[26-28]. In addition, the complex reflection coefficient from such a multilayer stack can be calculated by the transfer matrix method. Acting as fitted parameters, $Q_r$ and $Q_a$ are then determined by fitting calculated complex reflection coefficients using the transfer matrix method to Eq. (3).

The evolution of the fitted $Q_r$ and $Q_a$ as functions of PMMA thickness at a 35° incident angle (Figure 5f) is then systematically studied and depicted in Figure 5g. As the thickness of PMMA increases, both $Q_r$ and $Q_a$ values are simultaneously tuned. It is found that a positive topological charge, C=+1, is obtained when the system is tuned from overdamped ($Q_r<Q_a$) to underdamped ($Q_r>Q_a$), and vice versa[32]. In the overdamped process, the absorption rate is less than the radiation rate, while in the underdamped process, the absorption rate is greater than the radiation rate. It has previously been shown that by introducing loss into the system, topologically coherent perfect absorption charges with opposing values can emerge from the bound states in the continuum (BIC) while conserving topological charge[32]. Therefore, tuning the structure parameters tuning in this vertical cavity facilitates the competition between the radiation process and the absorption process, leading to the evolution of topological charge.

We clearly demonstrate that the enhanced light-matter coupling in MoS$_2$ stacked vertical cavity results in multiple perfect absorptions, leading to extraordinary levels of phase singularities. The generation, evolution, and annihilation of topological charges have been successfully demonstrated here. The concept of topological singularity is crucial in the study of topological radiative and scattering phenomena. The manipulation of singularities associated with perfect absorptions over a broad wavelength range will facilitate applications in polarization control, sensing, wireless power, and information transfer [14, 33].

## 3. Conclusion

In summary, unlike conventional distributed Bragg reflectors, multilayer planar FP cavity structures constructed by alternatively stacking dielectric PMMA layers and monolayer $MoS_2$ offer a free design parameter for controlling resonance modes and tuning light-matter interactions. The energy-momentum dispersions of the resonance modes are revealed through angle-resolved reflectance. The strong interaction between cavity photons and the monolayer $MoS_2$ leads to multiple perfect absorptions, resulting in topological singularity pairs, whose generation, evolution, and annihilation can be manually controlled. This expands the scope of various topological scattering phenomena by designing a vertical planar cavity structure. Our results promote the understanding of topological scattering photonics through resonance studies.

## 4. Methods

**Monolayer $MoS_2$ is grown using the CVD method:** Monolayer $MoS_2$ is grown using a molten salt method. A 5 mg/ml $Na_2MoO_4$ aqueous solution is prepared and spin-coated onto a $SiO_2$ (300 nm)/Si substrate at a rotation speed of 5000 rpm for 40 s. Before growth, high-purity argon gas is introduced into a quartz tube furnace at a flow rate of 80 sccm for 30 min. The spin-coated substrate is then placed on a ceramic boat facing upward and centered in the tube furnace. 50 mg of S powder is placed in another ceramic boat, positioned 20 cm upstream from the $Na_2MoO_4$ in the tube furnace. After CVD growth at 770°C for 8 min, the furnace is cooled down to room temperature over 120 min.

**Cavity structure fabrication:** A 34 nm layer of ZnO is deposited on the $SiO_2$(300 nm)/Si substrate using the atomic layer deposition (ALD) method. The ALD coating is carried out using an NCE-200R system (NanoFrontier, China) with water and diethylzinc as the oxygen and Zn precursors, respectively. The growth rate is 0.22 nm per cycle. $N_2$ is used as the process gas during the experiment. During the deposition, the Si/SiO$_2$ wafer is coated at 200 °C under a steady $N_2$ flow of 100 sccm, and each ALD cycle consisted of a 100 ms water and Diethylzinc pulse, followed by a 2.5 s purge with $N_2$. The Si/SiO$_2$ decorated with 34 nm ZnO is obtained after 160 ALD cycles. Subsequently, a wet transfer process is conducted to transfer MoS$_2$ flake onto the ZnO/SiO$_2$/Si substrate with the assistance of PMMA. First, PMMA is spin-coated onto the MoS$_2$ flakes to form a uniform film, followed by heat treatment on a hot plate at 80 °C for 2 min. Second, the PMMA coated MoS$_2$ flakes are soaked in a supersaturated NaOH solution for 12 h, which helps separate the MoS$_2$ flakes from the substrate, allowing them to float on the solution. Finally, the PMMA-supported MoS$_2$ flakes are transferred to the ZnO/SiO$_2$/Si substrate. By repeating the above steps multiple times, a vertical cavity structure with different periods can be fabricated.

**Angle-resolved reflectance measurements:** The angle-resolved reflectance of the vertical cavity with a period N=2 is measured using a NOVA-EX spectrometer (Shanghai Fu Xiang Ideaoptics. Inc) within an angular range of -60° to 60°, with a step size of 4°. The reflectance from the cavity structure is recorded in the wavelength range of 400 ~ 800 nm, with a step size of ~0.285 nm.

The phase difference for different polarized light in the vertical cavity structure with N=2 and a PMMA thickness of 500 nm is measured by an ME-L-L ellipsometer with a spot size of 50 μm (Wuhan Eoptics Technology Co). The ellipsometry parameters, $\psi$ and $\Delta\Phi$, are simultaneously measured in the wavelength range of 400 ~ 800 nm, with a step size is ~0.75 nm. Here, $\psi$ relates to the ratio of the reflection coefficient between p-polarized light and s-polarized light, while $\Delta\Phi$ represents the phase difference between p-polarized light and s-polarized light, as given by the formula $\frac{r_p}{r_s} = \tan(\psi)e^{i\Delta\Phi}$.

**Acknowledgements**

This work was supported the Research Foundation of Education Bureau of Hunan Province, China (Grant No. 18B477) and Dongguan Songshan Lake Introduction Program of Leading Innovative and Entrepreneurial Talents. The authors thank the helpful discussion with the engineers from Shanghai Fu Xiang Ideaoptics. Inc.

**Author contributions**

Y.W. and Z.S. conceived the idea, supervised the project, and revised the manuscript. Z.L. fabricated the microcavity samples. Z.L. and X.L. performed the experimental measurements and Z.L., Y.W. and Z.S. analysed it. Z.L. and Y.W. performed the theoretical modelling. B.Z. and W.L. provided discussions in the simulation. Z.L., Y.W. and Z.S. performed data analyzation and manuscript preparation. All authors contributed to the discussion of the results.

**Competing interests:**

The authors declare no competing financial interests.

**Data Availability Statement**

The data within this paper are available from the corresponding author upon reasonable request.